\documentclass[aps,prd,final,twocolumn,letterpaper, nofootinbib]{revtex4}
\usepackage{amsmath, amssymb, amsthm, cancel, comment, dsfont, enumerate, epstopdf, eucal, fancyhdr, graphicx, lipsum, mathtools, physics}
\usepackage{feynmp-auto}

\newcommand{\be}{\begin{equation}}
\newcommand{\ee}{\end{equation}}
\newcommand{\ba}{\begin{eqnarray}}
\newcommand{\ea}{\end{eqnarray}}
\newcommand{\bma}{\left(\begin{array}}
\newcommand{\ema}{\end{array}\right)}

\newcommand{\pont}{{\,^\ast\!}R\,R}

\begin{document}

\title{Chern-Simons Gravity and Neutrino Self-Interactions}
\author{Stephon Alexander}
\email{stephon_alexander@brown.edu}
\affiliation{Brown Theoretical Physics Center and Department of Physics, Brown University, 182 Hope Street, Providence, Rhode Island, 02903}
\author{Cyril Creque-Sarbinowski}
\email{creque@jhu.edu}
\affiliation{William H. Miller III Department of Physics and Astronomy, Johns Hopkins University, 3400 N. Charles St., Baltimore, Maryland 21218, USA}

\date{\today} 

\begin{abstract}
\noindent	
Dynamical Chern-Simons gravity (dCS) is a four-dimensional parity-violating extension of general relativity.  Current models predict the effect of this extension to be negligible due to large decay constants $f$ close to the scale of grand unified theories. Here, we present a construction of dCS  allowing for much smaller decay constants, ranging from sub-eV to Planck scales. Specifically, we show that if there exists a fermion species with strong self-interactions, the short-wavelength fermion modes form a  bound state. This bound state can then  undergo dynamical symmetry breaking and the resulting pseudoscalar develops Yukawa interactions with the remaining long-wavelength fermion modes. Due to this new interaction, loop corrections with gravitons then realize a linear coupling between the pseudoscalar and the gravitational Chern-Simons term. The strength of this coupling is set by the Yukawa coupling constant divided by the fermion mass. Therefore, since self-interacting fermions with small masses are ideal, we identify neutrinos as promising candidates. For example, if a neutrino has a mass $m_\nu \lesssim {\rm meV}$ and the Yukawa coupling is order unity, the dCS decay constant can be as small as  $f \sim 10^3 m_\nu \lesssim {\rm eV}$. We discuss other potential choices for fermions.  
\end{abstract}

\maketitle

\pagestyle{myheadings}
\markboth{Cyril Creque-Sarbinowski}{Chern-Simons Gravity and Neutrino Self-Interactions}
\thispagestyle{empty}

\section{Introduction}\label{sec:intro}
 The Einstein-Hilbert (EH) action is the most successful description of gravity today~\cite{1403.7377, 1501.07274}. However,  even at energies far below the regime of quantum gravity, there is no reason to expect the EH action to be a complete descriptor of gravity from an effective field theory (EFT) perspective~\cite{Cornwall:1974vz, Polchinski:1983gv}. A large body of work has thus been extensively researched in order to search for its deviations~\cite{Brans:1961sx, hep-th/0508049, 0805.1726, 1002.4928, 1106.2476, 1108.6266}. One extension is to look for parity-violation within our description of gravity, as we already know parity-violating effects to already be present within the Standard Model~\cite{Lee:1956qn, Glashow:1961tr}. The lowest-order term that encapsulates such parity violation is given by dynamical Chern-Simons gravity (dCS), where an additional pseudoscalar field $a$ is added that is linearly coupled to the Pontryagin density $^{\ast }RR$~\cite{gr-qc/0308071, 0907.2562}.

 dCS emerges naturally in the low-energy limit of string theory through the Green-Schwartz anomaly cancelling condition, with a decay constant that is typically the string scale \footnote{The scale of the decay constant can also depend on compactification volume factors.}~\cite{Green:1987sp, 0708.0001}. However, such large decay constants render compact astrophysical signals and superradiance effects to be undetectable. Attempts to generate dCS through radiative corrections by explicitly including Lorentz-violating terms~\cite{hep-th/0403205, 0708.3348, 0805.4409} have also been investigated, although it may be the case that such terms must vanish by gauge invariance~\cite{1903.10100}. Finally, dCS is also a possible solution to the gravitational analog to the strong CP problem in QCD if the pseudoscalar takes a suitable minimum~\cite{Deser:1980kc, hep-ph/9207208, 1808.01796}. In these cases, a gravitational axion is created either through the breaking of a global PQ~\cite{2108.05549, 2109.12920, Alexander:2021ssr} or axial~\cite{1602.03191, 1608.08969} symmetry.  
 
In this work, we show that dCS emerges as the low-energy limit of a theory of self-interacting fermions after dynamical symmetry breaking, i.e we do not require an additional scalar degree of freedom. In doing so, we also show that it is possible, but not necessary, to generate the mass of the fermion. We finish by demonstrating that the dCS decay constant can be made to reach the sub-eV regime for a wide range of phenomenological parameters. Specifically, we point out that light neutrinos are great candidates for reaching this threshold.   

This paper is organized as follows. First, in Sec.~\ref{sec:CSG}, we review the definition and dynamics of dCS. Then, in Sec.~\ref{sec:FSI}, we show how a single massive fermion with attractive self-interactions yields a bound state that undergoes spontaneous symmetry breaking. We specify the dynamics of the symmetry breaking in Sec.~\ref{sec:SSB} and show that it yields a Yukawa coupling between the bound state and fermion. With this new coupling, we then show, in Sec.~\ref{sec:Loop}, that it induces a gravitational Chern-Simons interaction through a triangle diagram with fermion loops. We then quantify the parameter space of generating dCS in Sec.~\ref{sec:param} and point out possible candidates for the fermion in Sec.~\ref{sec:fc}. We discuss and conclude in Sec.~\ref{sec:disc} and Sec.~\ref{sec:conc}, respectively.
\begin{fmffile}{simple_box}
\section{Chern-Simons Gravity}\label{sec:CSG}
Let $\kappa R$ be the EH term with $\kappa = (16\pi G)^{-1}$ and $R$ the Ricci scalar. Then the vacuum action $S$ of dCS gravity is given by  
\begin{align}
\label{eq:CSaction}
S =  \int d^4x \sqrt{-g} \left[\kappa R + \frac{a}{4f} \pont
- \frac{1}{2} \left(\nabla_\mu a\right) \left(\nabla^\mu a\right) \right].
\end{align}
The dCS extension beyond EH consists of the dynamical pseudoscalar field, $a$, which linearly couples  to the Pontryagin density through the dCS decay constant $f$. The Pontryagin density is defined as 
\be
\label{pontryagindef}
\pont={\,^\ast\!}R^\rho{}_\sigma{}^{\mu\nu} R^\sigma{}_{\rho \mu\nu}\,,
\ee
where 
\be
\label{Rdual}
{^\ast}R^\rho{}_\sigma{}^{\mu\nu}=\frac12 \epsilon^{\mu\nu\alpha\beta}R^\rho{}_{\sigma \alpha \beta}\,,
\ee
is Hodge dual to the Riemann tensor  and $\epsilon^{\mu\nu\alpha\beta}$ is the Levi-Civita tensor. The Pontryagin density can also be written in terms of divergence of the Chern-Simons topological current,   
\be
\nabla_\mu K^\mu = \frac14 \pont , 
\label{eq:curr1}
\ee
where 
\be
K^\mu :=\epsilon^{\mu\nu\alpha\beta}\left(\Gamma^\sigma{}_{\nu \rho}\partial_\alpha\Gamma^\rho{}_{\beta \sigma}+\frac23\Gamma^\sigma{}_{\nu \rho}\Gamma^\rho{}_{\alpha \lambda}\Gamma^\lambda{}_{\beta \sigma}\right)\,,
\label{eq:curr2}   
\ee
giving rise to the name ``Chern-Simons gravity''. Varying the action, Eq.~\eqref{eq:CSaction}, with respect to the metric yields the modified vacuum field equations, 
\begin{align}\label{eq:MetricEOM1}
G_{\mu\nu} + \frac{1}{\kappa f} \, C_{\mu\nu} &= \frac{1}{2 \kappa} T_{\mu\nu}\,,
\end{align}
which include a modification from the $C$-tensor, 
\begin{align}
C^{\mu\nu} = \left(\nabla_{\alpha} a \right) \; \epsilon^{\alpha \beta \gamma (\mu} \nabla_{\gamma} R^{\nu)}{}_{\beta} + \left(\nabla_{\alpha } \nabla_{\beta} a \right) \; {}^{\ast}R^{\beta(\mu\nu)\alpha}\,.
\label{eq:C-tensor}
\end{align}
The total energy-momentum tensor $T_{\mu\nu}$ is the sum of any matter stress-energy tensor (assumed in this subsection to be zero) and the stress-energy tensor of the pseudoscalar,
\begin{align}
T_{\mu\nu}^{(a)} = \left(\nabla_{\mu} a \right) \left( \nabla_{\nu} a \right) - \frac{1}{2} g_{\mu\nu} \left(\nabla_{\lambda} a \right) \left(\nabla^{\lambda} a \right)\,.
\end{align}
The pseudoscalar field itself obeys the following vacuum equation of motion
\begin{align}
\square a &= -\frac{1}{4 \kappa f} \, \pont\,,
\label{eq:theta-evolution}
\end{align}
which can be obtained by varying the action in Eq.~\eqref{eq:CSaction} with respect to $a$. 

dCS backreacted solutions of the pseudoscalar onto a gravitaional wave $h$ lead to an amplitude birefringence between left $h_L$ and right $h_R$ polarizations, resulting in parity violation:

\be h_{L,R} = e^{\pm A(a,\dot{a})_{L,R}}e^{ik^{\mu}x_{\mu}}. \ee

\begin{figure}

\begin{fmfgraph*}(150,100)
\fmfstraight
\fmfkeep{loop}
  \fmfleft{i}
  \fmfright{o1, o2}
  
  \fmf{dashes}{i,v1}
  \fmf{dbl_wiggly, tension = 0.5}{v2,o1}
  \fmf{dbl_wiggly, tension = 0.5}{v2, o2}
  \fmf{fermion,right,tension= 0.8}{v1,v2,v1}
\end{fmfgraph*}

\begin{fmfgraph*}(125,85)
\fmfstraight
    \fmfleft{i1,a,i2}
    
    \fmfright{o1,o2}
    \fmf{dbl_wiggly}{o1,t1}
    \fmf{dbl_wiggly}{t2,o2}
    \fmf{phantom,tension=0.5}{t1,i1}
    \fmf{phantom,tension=0.5}{t2,i2}
    \fmffreeze
    \fmf{fermion,tension=1}{t1,t2,t3,t1}
    \fmf{dashes,tension=2.0}{t3,a}
\end{fmfgraph*}


\caption{One-loop diagrams for the process $a \rightarrow hh$, with a pseudoscalar particle $a$ depicted as a dashed line and $h_{\mu\nu}$ the graviton as the solid wiggly line. The loops are generated by a fermion $\Psi$. }\label{fig:loops}
\end{figure}
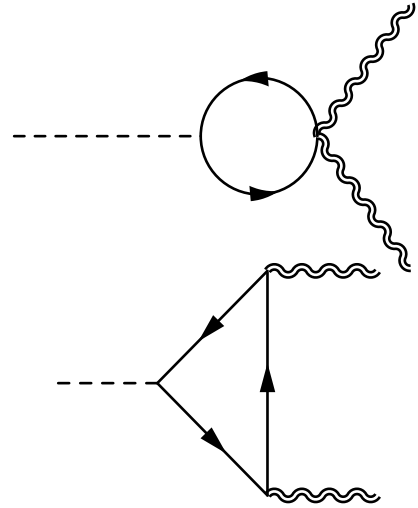

A key observation of dCS concerns the coupling $f$. Since $f$ has dimensions of energy, this theory is nonrenormalizable, seen at energies $E \gg f$, and so we are therefore motivated to search for a UV completion of dCS. More specifically, we expect that dCS action to be generated by loop diagrams of the UV complete theory, as shown by Fig.~\ref{fig:loops}.  In what follows, we show that fermions, and in particular neutrinos, that self-interact generate such diagrams. 

\section{Fermion Self-Interactions}\label{sec:FSI}
Our first goal is to derive a theory that undergoes spontaneous symmetry breaking from a fermionic, self-interacting theory. For simplicity, we consider the Lagrangian of a single massive fermion $\Psi$ with attractive self-interactions, 
\begin{align}
    \mathcal{L}_{\Psi} &= \bar{\Psi}\left(i\gamma^\mu\partial_\mu - \tilde{m}_\Psi\right)\Psi - \lambda \bar{\Psi}\Psi\bar{\Psi}\Psi, 
\end{align}
with $\gamma^\mu$ the Dirac gamma matrices, $\tilde{m}_\Psi$ the initial mass of the fermion, $\lambda$ the fermion self-interaction coupling constant, and interactions with gravitons currently suppressed. Similar to dCS, this Lagrangian is also nonrenormalizable, and thus has a cutoff $\Lambda$. Following the Nambu-Jones-Lasinio formalism presented in  \cite{Nambu:1961tp, Bardeen:1989ds, 0906.5161}, an attractive four-fermion interaction will generate a new bound state $\Phi$ below this cutoff. 

If we were to investigate this theory at extremely low energies $E \ll \Lambda$, where self-interactions dominate, we could perform a Hubbard-Stratonovich transformation on the fermion pair and integrate out the fermion from our theory, in accordance to the mean-field approximation.  Here, we consider the intermediate regime and perform a Hubbard-Stratonovich transformation on fermion modes between the scales $E$ and $\Lambda$, integrating them  out afterwards. More specifically, we first expand the fermion modes into long ($\ell,\ k < E$) and short ($s,\ k > E$ ) wavelength modes, $\Psi = \Psi_\ell + \Psi_s$. Next, we multiply the path integral of our theory by a constant,
\begin{align}
    Z_\alpha &= \int \mathcal{D}\alpha\mathcal{D}\bar{\alpha}\left(-\int d^4x\  \tilde{m}_\Phi^2 \bar{\alpha}\alpha\right),
\end{align}
with $\alpha$ an auxiliary field and $\tilde{m}_\Phi$ an as-of-yet unspecified bare mass scale. Then, we perform a field redefinition $\Phi = \alpha - \tilde{m}_\Phi^{-2}\bar{\Psi}_s\Psi_s$ in accordance with the Hubbard-Stratonovich transformation. We therefore get the new Lagrangian
\begin{align}
    \tilde{\mathcal{L}}_\Psi &= \bar{\Psi}(i \gamma^\mu \partial_\mu - \tilde{m}_\Psi)\Psi- \lambda\Big{[}\bar{\Psi}_\ell\Psi_\ell\bar{\Psi}_\ell\Psi_\ell \\
\nonumber    &+ 2\bar{\Psi}_\ell\Psi_\ell\bar{\Psi}_s\Psi_s + 2\bar{\Psi}_\ell\Psi_s\bar{\Psi}_\ell\Psi_s + 2 \bar{\Psi}_s\Psi_\ell\bar{\Psi}_s\Psi_\ell\Big{]} \\
\nonumber    &+ \tilde{m}_\Phi^{-4}\bar{\Psi}_s\Psi_s\bar{\Psi}_s\Psi_s +\bar{\Psi}_s\Psi_s \Phi + \Phi^*\bar{\Psi}_s\Psi_s + \tilde{m}_\Phi^2 \Phi^2.
\end{align}
Now, we integrate out the short scale modes and normalize the kinetic term,
\begin{align} \label{eq:effL}
       \tilde{\mathcal{L}}_\Psi &= \bar{\Psi}(i\gamma^\mu\partial_\mu - \tilde{m}_\Psi)\Psi - \lambda\bar{\Psi}\Psi\bar{\Psi}\Psi\\
\nonumber       &+ \left(\partial_\mu \Phi^* \right)\,\left( \partial^\mu \Phi \right) -
        y_\Phi\left(\Phi \bar{\Psi}\Psi + \text{h.c.}
        \right) \\
\nonumber &{}+
        m_\Phi^2 \, |\Phi|^2 -
        \frac{\lambda_\Phi}{4} |\Phi|^4,
\end{align}
with 
\begin{align}\label{eq:phi_params}
    m_\Phi^2 &= \left(\tilde{m}_\Phi^2 + \Pi_{0}\right)/\Pi_2,\\
    y_\Phi &= \lambda \Pi_{0}/\Pi_2^{1/2},\\
    \lambda_\Phi &= {\bf V}_4/\Pi_2^2,
\end{align}
renormalized parameters due to the condensation at one loop. Defining $\tilde{S}(k) = (-\cancel{k} + \tilde{m}_\Psi)/(k^2 + \tilde{m}_\Psi^2)$ to be Fourier transform of the fermion propagator,  
\begin{align}
    \Pi &= \int_{E}^\Lambda \frac{d^4\ell}{(2\pi)^4}\Tr\left[\tilde{S}(\ell)\tilde{S}(\ell + k)\right]\\
    &= -\frac{1}{16\pi^2}(\Lambda + E - 2k)(\Lambda - E)\\
    \nonumber &+ \frac{3\tilde{m}_\Psi^2}{16\pi^2}\log\left[\frac{\tilde{m}_\Psi^2 + (k + \Lambda)^2}{\tilde{m}_\Psi^2 + (k + E)^2}\right]\\
    \nonumber &+ \frac{\tilde{m}_\Psi^3}{8\pi^2 k}\left[\atan\left(\frac{E}{\tilde{m}_\Psi}\right) - \atan\left(\frac{\Lambda}{\tilde{m}_\Psi}\right)\right]\\
    \nonumber &+\left(\frac{\tilde{m}_\Psi^3}{8\pi^2 k} -\frac{3 \tilde{m}_\Psi k}{8\pi^2}\right)\left[\atan\left(\frac{k + \Lambda}{\tilde{m}_\Psi}\right) - \atan\left(\frac{k + E}{\tilde{m}_\Psi}\right)\right]\\
    \nonumber &+\frac{k^2}{16\pi^2}\log\left[\frac{\tilde{m}_\Psi^2 + (k + E)^2}{\tilde{m}_\Psi^2 + (k + \Lambda)^2}\right]\\
    &\equiv \Pi_{0} + k^2\Pi_{2},
\end{align}
 is then the self-energy and 
\begin{align}
{\bf V}_4 &= 6\int_E^\Lambda \frac{d^4\ell}{(2\pi)^4} \Tr\left[\tilde{S}(\ell)^4\right]\\
 &= \frac{6\tilde{m}_\Psi^2}{64\pi^2}\Bigg{[}\frac{45\tilde{m}_\Psi^4 + 119\tilde{m}_\Psi^2\Lambda^2 + 96\Lambda^4}{(\tilde{m}_\Psi^2 + \Lambda^2)^3}\\
 \nonumber&\ \ \ \ \ \ - \frac{22\tilde{m}_\Psi^4}{(\tilde{m}_\Psi^2  + E^2)^3} + \frac{73\tilde{m}_\Psi^2}{(\tilde{m}_\Psi^2 + E^2)^2} - \frac{96}{\tilde{m}_\Psi^2 + E^2}\Bigg{]}\\
 \nonumber &\ + \frac{24}{8\pi^2}\log\left[\frac{\tilde{m}_\Psi^2 + \Lambda^2}{\tilde{m}_\Psi^2 + E^2}\right],
\end{align}
the fermion-loop-corrected four-scalar vertex with zero external momenta. Note we have suppressed the long-wavelength subscript and ignore the renormalization of the original theory parameters for simplicity. Moreover, note that short-wavelength ($k\gg \Lambda$) modes are not dynamic as $\lim_{k/\Lambda \rightarrow \infty} \Pi_2 = 0$.

Each of the renormalized parameters has some running with energy,  and so in order to precisely determine the value of these parameters at some low-energy quasi-fixed point, a beta function analysis must be performed. Here, for simplicity and to capture the general scaling of parameters,  we assume the running of energy induces order one corrections to the parameters, i.e we take $\Pi_0 \sim -\Lambda^2/(16\pi^2)$ and $\Pi_2 \sim k^2/(16\pi^2)$. 

Restoring the ignored factors for a moment, we see that at zero external momentum, $\Pi_0$ simplifies to
\begin{align}
    \nonumber \Pi_0 &= -\frac{1}{16\pi^2}\Bigg{[}\left(\Lambda^2 - E^2\right) + 2\tilde{m}_\Psi^4\left(\frac{1}{\tilde{m}_\Psi^2 + E^2} - \frac{1}{\tilde{m}_\Psi^2 + \Lambda^2}\right)\\
     &- 3\tilde{m}_\Psi^2\log\left(\frac{\tilde{m}_\Psi^2 + \Lambda^2}{\tilde{m}_\Psi^2 + E^2}\right)\Bigg{]}.
\end{align}
Solving numerically, we find that $\Pi_0$ at $E = 0$ is negative as long as $\tilde{m}_\Psi \lesssim 0.74\Lambda$ and at any $E$ as long as $\tilde{m}_\Psi \lesssim 0.46\Lambda$~\footnote{However, given that factors of $\tilde{m}_\Psi/\Lambda$ are large in the opposite regimes, it is unclear if higher-order loops relax these conditions.}. Regardless, we infer the mass of $\Phi$ in Eq.~\eqref{eq:phi_params} can go from positive at $E \sim \Lambda$ to negative at $E \ll \Lambda$. Since there is a positive quartic term for $\Phi$, $\Phi$ can then undergo spontaneous symmetry breaking\footnote{As a matter of terminology, $\Phi$ undergoes spontaneous symmetry breaking, but $\Psi$ creates dynamical symmetry breaking.}.

\section{Spontaneous Symmetry Breaking}\label{sec:SSB}
 We now wish to realize the pseudoscalar coupling from spontaneous symmetry breaking, as this will be key to understanding the emergence of the gravitational Chern-Simons term.  Consider the following Lagrangian of a complex scalar field theory, 
\begin{align}\label{eq:phiL}
\nonumber \mathcal{L}_\Phi &= \bar{\Psi}_L\left(i\gamma^\mu\partial_\mu - \tilde{m}_\Psi\right)\Psi_L + \bar{\Psi}_R\left(i\gamma^\mu\partial_\mu - \tilde{m}_\Psi\right)\Psi_R\\
&+ \partial_\mu\Phi\partial^\mu \Phi^* 
- V\left(\left|\Phi\right|^2\right) - (y_\Phi \Phi \bar{\Psi}_L\Psi_R + {\rm h.c.}),
\end{align}
with $V(x)$ a $U(1)$ symmetry breaking potential and $y$ a new Yukawa coupling constant. Below some scale energy scale $F$, the field will fall to a minimum energy configuration that spontaneously breaks its $U(1)$ symmetry. As a result, we  expand the field around this minimum, $\Phi = (1/\sqrt{2})(F + \sigma)\exp[ia(x)/F]$. Plugging this expansion back into Eq.~\eqref{eq:phiL}, and also expanding the pseudoscalar exponential, we obtain
\begin{align}
\mathcal{L}_\Phi &= \bar{\Psi}\left[i\gamma^\mu\partial_\mu -\left(\tilde{m}_\Psi + \frac{y_\Phi F}{\sqrt{2}}\right)\right]\Psi\\
&\nonumber + \frac{1}{2}\partial_\mu a \partial^\mu a - \frac{y_\Phi}{\sqrt{2}}a\bar{\Psi}\gamma^5\Psi,
\end{align}
neglecting higher order terms of the pseudoscalar field. 

Typically, the symmetry breaking scale $F$ is related to the complex scalar's quadratic and quartic terms, $F = \sqrt{4|m_\Phi|^2/\lambda_\Phi}$, once $m_\Phi^2$ becomes negative. However, since the bare mass scale $\tilde{m}_\Phi$ in this phase is undetermined we cannot solve for $F$ using this expression. Therefore, in order to determine the symmetry breaking scale, or alternatively the mass $m_\Psi = \tilde{m}_\Psi + y_\Phi F/\sqrt{2}$, we invoke the self-consistency condition of the fermion propagator, i.e  the gap equation, depicted below,

\begin{eqnarray}\label{fig:gap_eq}
\nonumber \parbox{20mm}{\begin{fmfgraph*}(50,30)
    \fmfleft{i} 
    \fmfright{o} 
    \fmf{fermion}{i,o} 
    \end{fmfgraph*}} \quad & = \quad
\parbox{20mm}{\begin{fmfgraph*}(50,30)
    \fmfleft{i} 
    \fmfright{o} 
    \fmf{dashes_arrow}{i,o} 
    \end{fmfgraph*}} \quad + \quad
    \parbox{20mm}{\begin{fmfgraph*}(50,30)
    \fmfleft{i} 
    \fmfright{o} 
    \fmf{fermion}{i,v,v,o} 
    \fmfdot{v}
    \end{fmfgraph*}.}
\end{eqnarray}

The solid (dashed) arrow line is the total (bare) fermion propagator. The loop diagram is the result of the four-fermion interaction.

This condition amounts to a non-perturbative loop correction, $\Delta m_\Psi = y_\Phi F/\sqrt{2}$, to the fermion mass. Specifically, the gap equation relates the cutoff scale and self-interaction coupling constant $\lambda$ to this correction and the mass $m_\Psi$,
\begin{align}\label{eq:gapeq}
    \frac{\Delta m_\Psi}{m_\Psi} \left(\frac{2\pi^2}{\lambda \Lambda^2}\right) &= 1 - \frac{m_\Psi^2}{\Lambda^2}\log\left(1 + \frac{\Lambda^2}{m_\Psi^2}\right),
\end{align}
with a solution possible for sufficiently strong coupling, $\lambda \geq (\Delta m_\Psi/m_\Psi)(2\pi^2/\Lambda^2)$. For large cutoffs $\Lambda \gg m_\Psi$, this bound is saturated,
\begin{align}\label{eq:saturated}
\lambda &= \left(2.25\ {\rm MeV}\right)^{-2}\left(\frac{\Delta m_\Psi}{m_\Psi}\right)\left(\frac{10^{7}\ {\rm eV}}{\Lambda}\right)^2,
\end{align}
while for small cutoffs $\Lambda \ll m_\Psi$
\begin{align}\label{eq:small_cutoff}
\lambda &= \left(1.59\ {\rm MeV}\right)^{-2}\left(\frac{\Delta m_\Psi}{m_\Psi}\right)\left(\frac{m_\Psi^2}{\Lambda^2}\right)\left(\frac{10^7\ {\rm eV}}{\Lambda}\right)^2.
\end{align}
Solving for $F$, and ignoring the lower-order terms in $y_\Phi$, we find that 
\begin{align}
    F &\approx m_\Psi\frac{2\sqrt{2}}{\pi}\left[1 - \frac{m_\Psi^2}{\Lambda^2}\log\left(1 + \frac{\Lambda^2}{m_\Psi^2}\right)\right].
\end{align}
Therefore, for large cutoffs, $F \approx 0.9 m_\Psi$ and for small cutoffs, $F \approx 0.45 \Lambda^2/m_\Psi$. Moreover, we point out that at low cutoffs, the condition $\tilde{m}_\Phi \lesssim c\Lambda,\ c\sim 1$ translates to ($\Delta m_\Psi/m_\Psi) \gtrsim 1 - c(\Lambda/m_\Psi)$ or $\Delta m_\Psi/m_\Psi \sim 1$.

Thus, through this mechanism, we are able to generate a Yukawa coupling between the fermion and a pseudoscalar.

\section{Loop Generation of dCS}\label{sec:Loop}
We now place our fermion in a curved background in order to realize dCS. That is, we begin with a Dirac fermion $\Psi$ that has both a minimal gravity and a pseudoscalar Yukawa interaction,
\begin{align}\label{eq:psiL}
\mathcal{L}_g &= \bar{\Psi}\left(i\gamma^b e_b^\mu D_\mu - m_\Psi\right)\Psi + iga\bar{\Psi}\gamma^5\Psi,
\end{align}
with  $e^\mu_b$ an orthonormal tetrad basis, $D_\mu = \partial_\mu - (i/4)\sigma^{ab} \omega_{\mu ab}$ the covariant spinor derivative, $\sigma_{ab} = [\gamma^a, \gamma^b]$ the Lorentz group generator for Dirac fermions, $\omega_{\mu ab} = (1/2)e_a^\nu(\partial_\mu e_{b\nu} - \partial_\nu e_{b\mu}) - (1/2)e_b^\nu(\partial_\mu e_{a\nu} - \partial_\nu e_{a\mu}) + (1/2)e_a^\rho e_b^\sigma (\partial_\sigma e_{c\rho} - \partial_\rho e_{c\sigma})e_\mu^c$ the torsion-free spin connection, and $g$ the pseudoscalar Yukawa coupling constant (equal to $y_\Phi/\sqrt{2}$ in the previous section). Here, Greek indices indicate the global Lorentzian structure and Latin indices the local structure.   

In order to generate dCS, we perform the same steps as to derive the pion-photon decay and integrate out the fermion triangle loop from our theory. For energies below the fermion mass $m_\Psi$, integrating out the field $\Psi$ is equivalent to evaluating the $\Psi$-dependent Lagrangian in a fixed gravitational field. As a result, the effective Lagrangian  is then
\begin{align}
\mathcal{L}^{\rm eff}_g &= g a J_a, \\
J_a &= i\bra{h}\bar{\Psi}\gamma^5\Psi\ket{h},
\end{align}
with $J_a$ is the pseudoscalar composite operator, expressed as an expectation value over a gravitational field $h$. In order to evaluate this expectation value, we use the alternative, and equivalent, expression for the gravitational Adler-Bell-Jackiw (ABJ) anomaly,
\begin{align}\label{eq:gABJ}
\bra{h}\partial_\mu \tilde{J}_5^{\mu}\ket{h} &= -\frac{1}{384\pi^2}{^\ast}RR,
\end{align} 
with $\partial_\mu \tilde{J}_5^{\mu} = 2m_\Psi i\bar{\Psi}\gamma^5\Psi$ the divergence of the classical axial current associated with $\Psi$. Note that this current is not to be confused with the total axial current $\partial_\mu J_5^\mu = \partial_\mu \tilde{J}_5^\mu + {^\ast}RR/(384\pi^2)$. The expression for the ABJ anomaly can be obtained in numerous different ways (e.g. Fujikawa's method)~\cite{Delbourgo:1972xb, Alvarez-Gaume:1983ihn, Fujikawa:2004cx}. The equivalent perturbative triangle loop diagram results in inserting the interaction vertex between the graviton and the fermion and evaluating the amplitude integrals. The relevant graviton-fermion interactions for the amplitude can be found in Ref.~\cite{gr-qc/0607045}. For example, a typical triangle amplitude for the decay of the pseudoscalar to gravitons is:
\begin{align} 
\mathcal{M}(a \rightarrow hh) &=\frac{ga}{16}\int d^{4}k \epsilon^{\mu\nu}(q_{1})\epsilon^{\delta\gamma}(q_{2})\\
\nonumber&\times\left[ \tilde{S}(k-q_{1})_{\gamma}\gamma_{\delta}\tilde{S}(k-q_{1})_{\nu}\gamma_{\mu}\gamma_{5} \right]\\
\nonumber &+  (\mu\nu \rightarrow \delta\gamma ;\ 1 \rightarrow 2)
\end{align}
where $\epsilon^{\mu\nu}$ is the graviton polarization tensor.
However, since this diagram is one-loop exact, the choice of computational scheme does not alter the result. Therefore, since $J_a = [1/(2m_\Psi)]\bra{h}\partial_\mu \tilde{J}_5^{\mu}\ket{h}$, we obtain the CS term
\begin{align}\label{eq:psiCSG}
\mathcal{L}^{\rm eff}_{g} &= -\frac{g}{384\pi^2}\frac{a}{2m_\Psi}{^\ast}RR.
\end{align} 
In this procedure, CSG is specified by two parameters: the mass of the fermion, and its Yukawa coupling to the pseudoscalar $a$. Moreover, connecting Eq.~\eqref{eq:psiCSG} with Eq.~\eqref{eq:CSaction} we see that the dCS decay constant is
\begin{align}\label{eq:dCS_dc}
f = 192\pi^2 \frac{m_\Psi}{g}.
\end{align}
Therefore, when the cutoff is large, $\Lambda \gg m_\Psi$, we have
\begin{align}\label{eq:min_dc}
f &= 1.7\ {\rm eV}\ \left(\frac{\Delta m_\Psi}{ m_\Psi}\right)^{-1}\left(\frac{m_\Psi}{10^{-3}\ {\rm eV}}\right),
\end{align}
and when it is small
\begin{align}\label{eq:max_dc}
f &= 0.85\ {\rm eV}\ \left(\frac{\Delta m_\Psi}{ m_\Psi}\right)^{-1}\left(\frac{\Lambda}{m_\Psi}\right)\left(\frac{\Lambda}{10^{-3}\ {\rm eV}}\right).
\end{align}
Thus, if a fermion self-interacts strongly enough, Chern-Simons gravity is a consequence. 
\section{Parameter Space}\label{sec:param}
We now address the relevant parameter space for observations. First, we consider the most general parameterization to generate dCS with a four-fermion interaction, and then consider two specific examples. 
\subsection{General}
\begin{figure}
    \includegraphics[width = \linewidth]{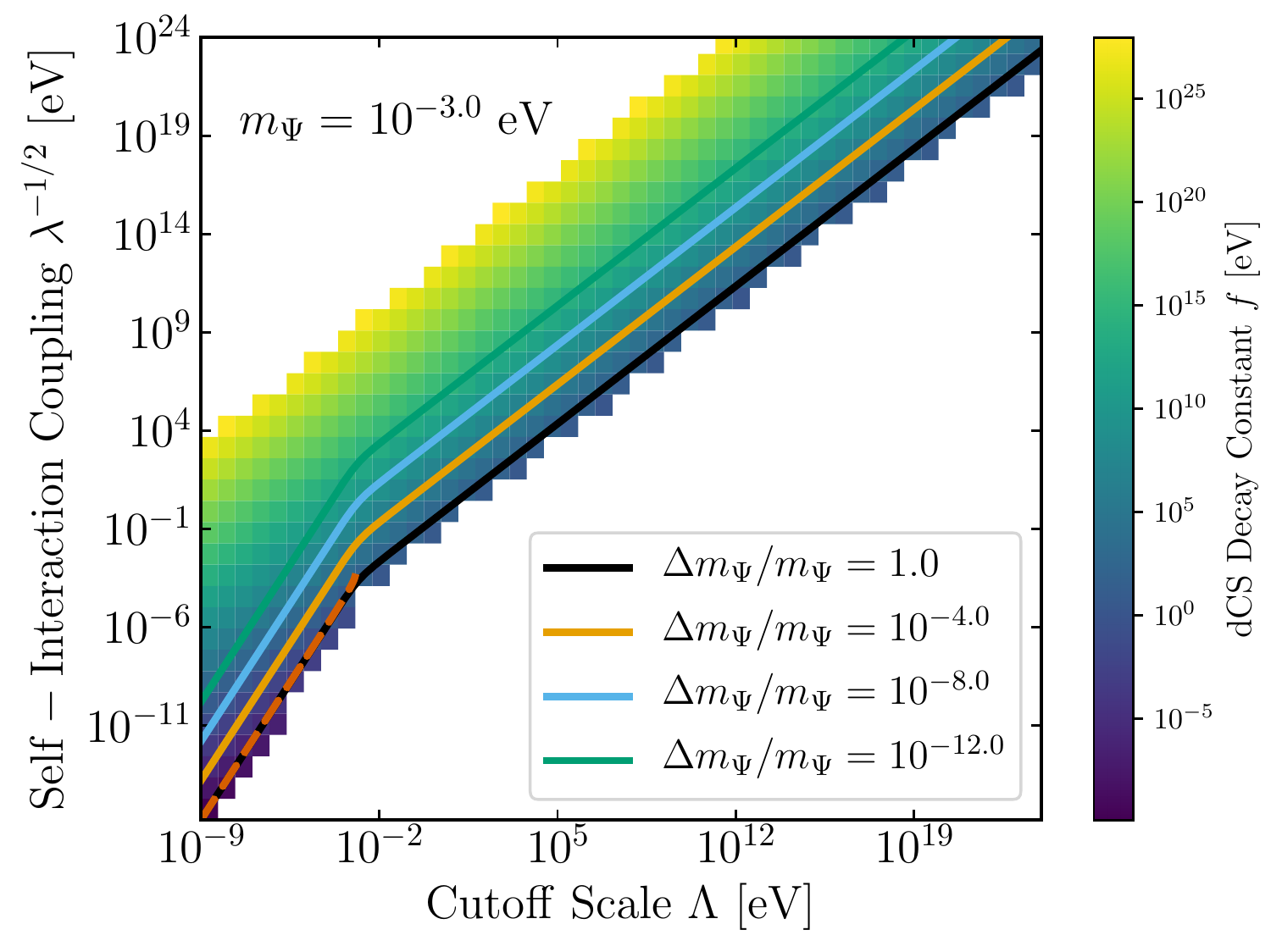}
    \caption{The parameter space for dCS generated by a self-interacting fermion $\Psi$ of mass $m_\Psi = 10^{-3}\ {\rm eV}$ using Eq.~\eqref{eq:gapeq} and Eq.~\eqref{eq:dCS_dc}. The black (orange) [blue] \{green\} solid line indicates the parameters necessary to generate a mass correction $\Delta m_\Psi/m_\Psi = 1\ (10^{-4})\  [10^{-8}]\  \{10^{-12}\}$. The dashed red line indicates the regime when the condition $\tilde{m}_\Psi \lesssim 0.74\Lambda$ is active. Regions strictly above this red line do not undergo spontaneous symmetry breaking (SSB), and thus cannot generate dCS,  if the one-loop calculation is valid (the red line itself can undergo SSB). Finally, we only plot parameters that  are below the Planck scale, and have $\Delta m_\Psi/m_\Psi \leq 1$. Moreover, while in principle the cutoff can be made arbitrarily small, we only plot $\Lambda \gtrsim 10^{-6}m_\Psi$ for visualization purposes.} 
    \label{fig:Lvl}
\end{figure}

\begin{figure}
    \includegraphics[width = \linewidth]{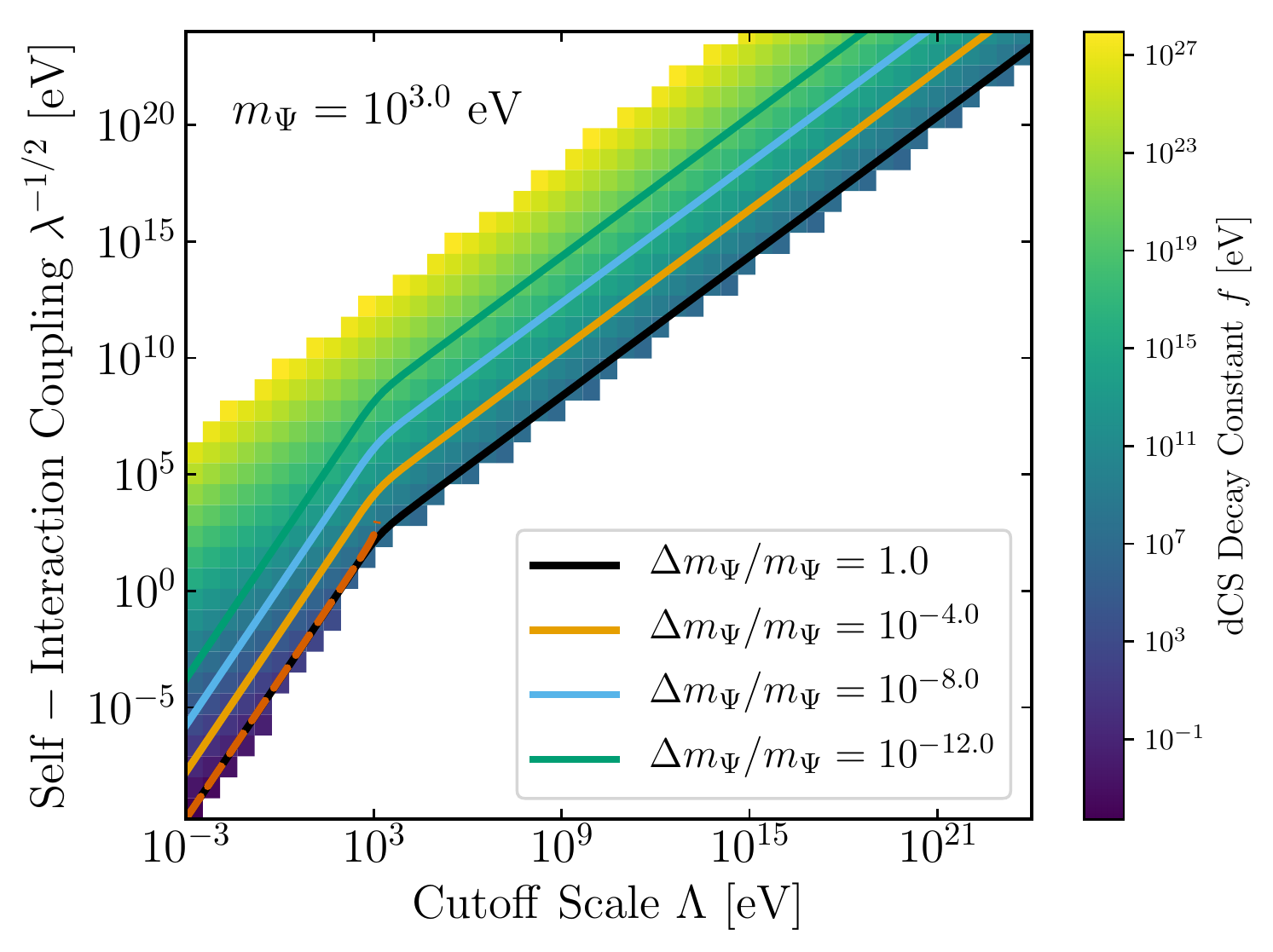}
    \caption{Same as Fig.~\ref{fig:Lvl}, except for $m_\Psi = 10^3\ {\rm eV}$.} 
    \label{fig:Lvl_2}
\end{figure}

 From the gap equation [Eq.~\eqref{eq:gapeq}], the main parameters in question are the mass of the fermion, $m_\Psi$, the cutoff scale $\Lambda$, and the self-interaction coupling strength $\lambda$. We plot the relation between these quantities to the dCS decay constant $f$ in Fig.~\ref{fig:Lvl} and Fig.~\ref{fig:Lvl_2}.

\subsection{Scalar Mediator}
One scenario to create a fermion self-interaction is through the interaction with a scalar mediator $\chi$ of mass $m_\chi$. The interaction has a Yukawa-like form, $\mathcal{L} \supset g_\chi\chi \bar{\Psi}\Psi$. At energies $E \ll m_\chi$, a four-fermion interaction is induced with $\lambda = (g_\chi/m_\chi)^2$, indicating the cutoff of the theory is roughly $\Lambda \sim m_\chi$. 

As a result, the gap equation now takes the form
\begin{align}
\frac{\Delta m_\Psi}{m_\Psi} \frac{2\pi^2}{g_\chi^2} = 1 - \frac{m_\Psi^2}{m_\chi^2}\log\left(1 + \frac{m_\chi^2}{m_\Psi^2}\right).
\end{align}
Therefore, for $m_\chi \gg m_\Psi$, 
\begin{align}
g_\chi^2 &= 2\pi^2\frac{\Delta m_\Psi}{m_\Psi},
\end{align}
and for $m_\chi \ll m_\Psi$,
\begin{align}
g_\chi^2 &= 4\pi^2\frac{m_\Psi^2}{m_\chi^2}\frac{\Delta m_\Psi}{m_\Psi}.
\end{align}
Moreover, for both regimes,
\begin{align}
f &= 34\ {\rm eV}\ g_\chi^{-2}\ \left(\frac{m_\Psi}{10^{-3}\ {\rm eV}}\right).
\end{align}
\subsection{Gravitational Torsion}
If the gravitational connection has non-zero torsion, a scalar four-fermion interaction is typically induced at energies below the Planck scale for all fermions~\cite{hep-th/0507253}. In this case, the self-interaction coupling constant is $\lambda = 3\pi/\Lambda_T^2$. 
\section{Fermion Candidates}\label{sec:fc}
\subsection{Neutrinos}
Since smaller fermion masses yield larger coupling constants, we are most interested in fermions with small masses that self-interact. The ideal candidate for such a scenario is thus the neutrino.  While the Standard Model induces neutrino self-interactions, these are too small to realize a condensate in nature~\cite{hep-th/0407164}. However, beyond the Standard Model interactions are both a viable option and well-motivated~\cite{2203.01955}. 

In regards to the parameter space, oscillation experiments~\cite{1811.05487}, along with Planck 2018 measurements~\cite{1807.06209}, imply that an active neutrino's mass can be as small as $m_\nu \sim {\rm meV}$ for the normal hierarchy, or even be massless in the inverted hierarchy~\cite{2005.05332}. In addition, sterile neutrinos may have masses as small as any given active neutrino, and as large as the Planck mass~\cite{hep-ph/9303287, 1204.5379, 1303.6912, 1609.07647, hep-ph/0006358, 2203.08039, 2206.01140}. Given the wide range of masses, connecting our results to specific neutrino self-interaction models (and therefore giving constraints on $\lambda$  and $\Lambda$) is beyond the scope of our work.

\subsection{Fermionic Self-Interacting Dark Matter}
 It is also possible that the fermion in question is a dark matter (DM) particle that self-interacts~\cite{1012.5317, 1705.02327, 2110.15391, 2112.09057, 2206.14395}.  In this case, if this particle compromises all of DM, it may have masses between $0.1\ {\rm GeV} \leq m_\Psi \leq 10^7\ {\rm GeV}$~\cite{1310.3509}. However, if instead the fermion is one of a large number $N \lesssim 10^{62}$ of particles, such as in ultralight fermionic dark matter~\cite{2008.06505}, the mass of the fermion could possibly be as small as $m_\Phi \sim 10^{-14}\ {\rm eV}$. We note, however, that the ultralight case with self-interactions has not been studied, and so it is not definite that such a scenario is viable.

\section{Discussion}\label{sec:disc}
We clarify three assumptions and give four comments. First, given that we are dealing with an EFT with a very low energy cutoff, one may worry that astrophysical or cosmological systems, such as compact binary coalescences, are characterized by larger energies. As a rough argument, the scale of the EFT breaking down at ringdown occurs at energies smaller than the Schwarzschild radius, which for $\sim $eV energies is far above the EFT limit. A much more detailed analysis of such scales has been done in Ref.~\cite{1810.07706, 2111.02072}, although they note that it is difficult to analyze the parity-violating sector. 

Second, for simplicity, we only considered the self-interaction of a single neutrino. Our method of generating a neutrino bound state can be extended to multiple generations of neutrinos in straightforward fashion, whether they be active or sterile, through  promoting the self-interacting coupling constant to a self-interacting coupling matrix. 

Third, we assume that renormalization will lead to order one changes in the induced condensation parameters. We base this assumption on both dimensional grounds and the results presented in other dynamical symmetry breaking papers~\cite{Bardeen:1989ds, hep-ph/0211385}. However, we note that as energies fall, the Yukawa coupling $y_\Phi$ increases. Therefore, the dCS decay constant will decrease and our results are then in fact upper bounds on the actual decay constant and the likelihood of detecting dCS via this mechanism increases. 

It may be the case that the complex scalar and fermion in Sec.~\ref{sec:SSB} are actually fundamental new particles, rather than related to some bound state. In this case, the dCS phenomenology is completely specified by the Yukawa coupling and symmetry breaking scale. 

If the fermion in question is a neutrino, then the pseudoscalar interaction in Eq.~\eqref{eq:psiL} can also be generated through the breaking of a global lepton symmetry, so that $a = J$, with $J$ the majoron~\cite{Kim:1986ax, Gelmini:1980re, Chikashige:1980ui, 1709.07670}. Moreover, this association is also possible if the fermion is a dark-matter candidate~\cite{1408.4929, 2105.04255},  however the masses of the fermion typically are large $m_\Psi \gg {\rm GeV}$. 

In order to create fermion self-interactions we considered the case where a real scalar mediator generates the fermion self-interactions, however it can equally be a complex scalar, vector, or tensor. For these mediators, we expect the resulting low-energy self-interacting coupling constant to be an $\mathcal{O}(1)$ factor difference from the real scalar mediator. 

Finally, we point out that sensitivity forecasts of 2G detectors to the dCS decay constant in the inspiral of black hole mergers  yield $f \gtrsim 10^{-50}\ {\rm eV}$ (or $\xi^{1/4} \lesssim 10\ {\rm km}$)~\cite{1208.5102, 1712.00682, 2205.02675}. Hence, it will be difficult, but not implausible, that inspiral signals will yield a detectable signal from our mechanism. In particular, it may be the case that backreaction of the pseudoscalar onto the evolution of the binary could yield a more distinct imprint on gravitational wave observations. In addition, it is unclear if merger and ringdown signals will yield larger parity-violating signals, given that the scale of the system is much smaller and the gravitational strength much larger, especially in supermassive black hole systems. We leave the investigation of all these possibilities for future work. 

\section{Conclusion}\label{sec:conc}
In this paper, we presented a new method of generating dCS, with decay constants that reach the sub-eV regime. First, we showed that a fermion with attractive self-interactions will form a bound state made up of short-wavelength fermion modes. In doing so, these modes also generate quadratic and quartic potential terms for the bound state, allowing for spontaneous symmetry breaking to occur.

After demonstrating symmetry breaking, we then showed that the angular part of the bound state corresponds to a pseudoscalar that developed Yukawa interactions with the long-wavelength fermion modes.  In order to solve for the symmetry breaking scale, we invoked the gap equation, a self-consistency condition for the fermion propagator. As a result, we found that the symmetry breaking scale at large cutoffs is roughly the scale of the fermion mass and at small scales it is roughly the square of cutoff scale divided by the fermion mass. 

The Yukawa interaction then facilitated a triangle diagram between the pseudoscalar and a pair of gravitons, with the long-wavelength fermions going around the loop. By integrating out these fermions, we then saw the dCS interaction term was created, with a decay constant that is proportional to the mass of the fermion divided by the Yukawa coupling. 

With these results, we then identified neutrinos as particularly optimistic candidates to generate dCS, due to their small masses and ability to self-interact. In particular, we found that for neutrinos of mass $m_\nu\lesssim\ {\rm meV}$ the dCS decay constant could be less than an eV, $f \sim 10^3m_\nu \lesssim {\rm eV}$. 
\end{fmffile}
\subsection*{Acknowledgments}
 We especially thank Marc Kamionkowski for discussions, criticisms, and encouragements on this paper. We thank Peter Adshead for useful discussions. S.A thanks Glennys Farrar for pointing out the possibility that the dCS decay constant could reach the neutrino mass scale.  C.C.S would like to thank Robert Jones for useful conversations about condensation and the Aspen Center for Physics, which is supported by National Science Foundation grant PHY-1607611, for hospitality while this work was completed. C.C.S also acknowledges the support of the Bill and Melinda Gates Foundation. This work was supported by the Simons Foundation.

\end{document}